# Dual Enhancement of Superconductivity in $FeSe/SrTiO_3$ via Orbital and Correlation Synergy

Guihao Jia[a,b,1], Jingming Yan[a,b,1], Yucong Peng[a,b], Shendong Su[a,b], Pei Ouyang[a,b], Xiaopeng Hu[a], Qi-Kun Xue[a,b,c,d,e,*], and Wei Li[a,b,e,*]

[a]State Key Laboratory of Low-Dimensional Quantum Physics, Department of Physics, Tsinghua University, Beijing 100084, China; [b]Frontier Science Center for Quantum Information, Beijing 100084, China; [c]Beijing Academy of Quantum Information Sciences, Beijing 100193, China; [d]Southern University of Science and Technology, Shenzhen 518055, China; [e]Hefei National Laboratory, Hefei 230088, China

[1]G.J. and J.Y. contributed equally to this work.

*Corresponding authors: Qi-Kun Xue, and Wei Li.

**Email:** qkxue@mail.tsinghua.edu.cn, weili83@tsinghua.edu.cn

**Author Contributions:** W.L. and Q.-K.X. designed research; G.J., J.Y., Y.P., S.S., and P.O. performed research; W.L. and Q.-K.X. contributed new reagents/analytic tools; G.J., J.Y., Y.P., S.S., and W.L. analyzed data; G.J. and W.L. wrote the paper.

**Competing Interest Statement:** The authors declare no competing interests.

**Abstract**

In iron-based superconductors, the $d_{z^2}$ orbital band typically resides far below the Fermi level and has not been considered to participate in Cooper pairing. Here, using monolayer $FeSe/SrTiO_3$ as a model system, we demonstrate that tip-induced tensile strain controllably shifts the $d_{z^2}$ band toward the Fermi level, driving a two-stage enhancement of superconductivity. In-plane lattice expansion first enhances electronic correlation, amplifying superconductivity in the initial stage. As strain further increases, the upward-shifted $d_{z^2}$ band hybridizes with the $d_{xy}$ band, reconstructing the pairing-active *d*-orbital bands and inducing a secondary, stronger gap enhancement. Collectively, these two stages enlarge the superconducting gap from 17.8 to 23.6 meV. Throughout this process, invariant Fermi wave vectors confirm that the enhancement originates from band renormalization and reconstruction rather than carrier doping. Our work establishes a route to tailor superconducting states via strain-activated electronic correlations and band engineering, and reveals a previously unrecognized orbital-selective pairing mechanism with broad implications for correlated multiband superconductors.

**Significance Statement**

In iron-based superconductors, the $d_{z^2}$ orbital band lies deep below the Fermi energy and is conventionally excluded from the superconducting pairing process. This work overturns this view. Using a scanning tunneling microscope tip to apply local mechanical strain, we dynamically shift $d_{z^2}$ band upward, activating its direct role in Cooper pairing and enhancing the superconducting gap by over 30%. This strain-driven mechanism, which operates independently of chemical doping, establishes a new paradigm for controlling superconductivity via orbital-selective interactions and offers a general strategy for designing improved superconducting materials.

**Main Text**

**Introduction**

The discovery of enhanced superconductivity in monolayer $FeSe/SrTiO_3$ (FeSe/STO) (1) has sparked transformative interest in the field of high-temperature superconductivity (2-13). This enhancement is attributed to two types of interfacial interactions between FeSe and STO (3-5). Specifically, optimal doping is inherently achieved through charge transfer from STO substrate (3-5), and the interfacial electron-phonon coupling contributes additionally (2,5,6,9,11,12). However, both interactions face limitations: (i) additional carrier doping cannot further amplify the superconductivity, as excess electrons or holes suppress the superconducting gap (14,15). (ii) Strengthening the electron-phonon coupling is constrained by the substrate's intrinsic properties and epitaxial growth kinetics of FeSe (16).

Electronic correlation offers an alternative pathway to overcome these limitations and further amplify superconductivity in FeSe/STO. Strong electron correlations and symmetry-breaking electronic orderings have been observed in both cuprate superconductors (17-19) and iron-based superconductors (20-24). Strong electronic correlations are manifested through a stripe-like smectic phase in FeSe/STO (25,26). This smectic ordering becomes more pronounced in multilayer $FeSe/BaTiO_3$ (27), where both lattice constant and electronic correlation exceed those in FeSe/STO. Concurrently, superconducting gaps of monolayer and electron-doped multilayer $FeSe/BaTiO_3$ surpass their FeSe/STO counterparts (8,16,27). These results together reveal that in-plane lattice expansion strengthens electronic correlations, thereby providing an additional dimension to optimize superconductivity (26,27).

In iron-based superconductors, the $d_{z^2}$ band lies deeply below the Fermi level ($E_F$) and is generally regarded as irrelevant to the Cooper pairing mechanism (28-30). However, this view may overlook its potential role, a gap our work aims to address by providing direct evidence of its active contribution. Moreover, the energy of the $d_{z^2}$ band in FeSe can serve as a reliable indicator of electronic correlations via their shared sensitivity to lattice constants (30-34). Upon in-plane lattice expansion and out-of-plane compression (see crystal structure in Fig. 1*A*), electronic correlations are strengthened by lengthening the dominant hopping path, concurrently shifting the $d_{z^2}$ band toward $E_F$. This rigorously established band shift (4,16,30-34) directly reflects the electronic correlation strength.

Using scanning tunneling microscopy (STM), we demonstrate a two-stage process that further enhances the superconductivity through synergy of electronic correlations and $d_{z^2}$ band engineering in monolayer FeSe/STO. Sub-ångström-precision tip approach induces in-plane lattice expansion in FeSe via repulsive tip-sample interaction, and the resulting enhanced electronic correlations, as evidenced by the $d_{z^2}$ band shift toward $E_F$, constitute the first stage of gap enhancement. In the second stage, a secondary stronger gap amplification emerges when the $d_{z^2}$ band hybridizes with the flat $d_{xy}$ band at −80 meV, directly linking the *d*-orbital band reconstruction to superconducting gap enhancement. This work provides the first experimental demonstration of $d_{z^2}$ band-driven gap amplification, revealing its influence in Cooper pairing and necessitating a revised theoretical framework for pairing mechanism.

## Results

### Tip-driven dual enhancement of superconducting gaps

Figure 1*B* schematically depicts our STM experimental configuration. Based on the typical dependence of tunneling current on junction width in STM principle (35), the tip-sample distance is precisely controlled by adjusting the current at a fixed bias under feedback. Subsequent scanning tunneling spectroscopy (STS, or d*I*/d*V* spectrum) measurements are performed at stabilized tip-sample distances (defined by setpoint bias and current) by sweeping bias with the feedback loop disabled.

High quality monolayer FeSe film with an atomically flat and homogeneous surface morphology is prepared on the STO substrate by molecular beam epitaxy (MBE) growth (Fig. 1 *C*-*D*). U-shaped superconducting gap with characteristic double coherence peaks are observed (100 pA curve in Fig. 1*E*) on the prepared sample, consistent with previous reports (1). Intriguingly, with increasing tunneling current (i.e., reducing tip-sample distance), the coherence peaks shift anomalously away from $E_F$, resulting in enlargement of superconducting gaps (dashed guidelines in Fig. 1*E*). This gap amplification is quantitatively demonstrated through systematically acquired d*I*/d*V* spectra across varying currents (Fig. 1*F*), which not only track the continuous energy shift of coherence peaks but also reveal an obvious transition during the amplification process (see purple arrow in Fig. 1*F*). The corresponding second-derivative intensity map of the d*I*/d*V* spectra [$-d^2(dI/dV)/dV^2$] further highlights the evolution of coherence peaks (red-colored features marked by arrows in Fig. 1*G*) with logarithmic current scale.

Figure 1*H* quantitatively displays the current-dependent evolution of superconducting gaps on a logarithmic current scale. From 100 pA to 9 nA, the outer gap ($\Delta_1$) rises from 17.8 meV to 23.6 meV (32% increase), while the inner gap ($\Delta_2$) increases from 11.5 meV to 16.0 meV (39% enhancement). Both gaps exhibit a two-stage evolution. Below 1.25 nA, $\Delta_1$ rises modestly from 17.8 to 19.7 meV (10%), and $\Delta_2$ increases marginally from 11.5 to 12.2 meV (6%). Above 1.25 nA, both gaps undergo synchronous rapid enhancement, contributing the remaining gap increases (22% for $\Delta_1$ and 33% for $\Delta_2$). Notably, the slopes of both gaps sharpen abruptly around 1.25 nA (dashed lines in Fig. 1*H*), suggesting the emergence of a critical transition that activates the stronger secondary superconducting gap amplification. This transition region, defined by the intersection of the two-stage evolution trends (dashed lines in Fig. 1*H*), is marked by purple arrows in Fig. 1 *F* and *H* and a purple dashed line in Fig. 1*G*, indicating the onset of the secondary superconducting gap amplification. This gap evolution has been repeated across multiple samples and tips cross-validated in two independent STM-MBE systems (see *SI Appendix*, Fig. S1). Moreover, the U-shaped superconducting gaps persist during the superconducting gap amplification, and the density-of-states (DOS) at $E_F$ remain zero (*SI Appendix*, Fig. S2).

***d*-orbital band reconstruction and superconducting gap dual amplification**

To investigate the origin of the gap amplification, we systematically measured the current-dependent electronic structure. Figure 2*A* presents a typical wide-bias-range d*I*/d*V* spectrum of monolayer FeSe/STO acquired under conditions that minimize tip perturbation (see *SI Appendix*, Fig. S3), consistent with previous studies (36). The d*I*/d*V* spectrum measures the local DOSs, and we assign the observed peaks and kinks in Fig. 2*A* to specific *d*-orbital band edges (Fig. 2*B*, see *SI Appendix*, Note S1 and Fig. S4 for details) by comparing them with the band structure determined by angle-resolved photoemission spectroscopy (ARPES) (5,11,16,32,37,38). Alternative interpretations, such as inelastic electron tunneling processes mediated by bosonic modes, are excluded (*SI Appendix*, Note S1). The suppression of spectral weight near $E_F$ in the d*I*/d*V* spectrum arises from the momentum-space integration inherent to STS, which yields a dominant contribution from states around the $\Gamma$ point (7) (*SI Appendix*, Fig. S4).

Figure 2*C* displays d*I*/d*V* spectra measured over a wide range of setpoint current from 5 pA to 6 nA, spanning three orders of magnitude at a fixed setpoint bias voltage of 30 mV (corresponding junction resistances from 6 GΩ to 5 MΩ). The measurement varies about 260 picometers of tip-sample distance reduction (lower inset of Fig. 2*C*). The spectrum at 5 pA, corresponding to the farthest tip-sample distance in that series to maintain a stable tunneling junction, represents the pristine band structure of unperturbed FeSe/STO, consistent with the spectrum in Fig. 2*A*. The current-dependent waterfall plot delineates the concurrent evolutions of *d*-orbital bands and superconducting gaps, with dashed lines (color-matched to arrows) tracking the band energy shifts. As the STM tip approaches the surface, the $d_{xz/yz}$ and $d_{xy}$ band kinks retreat slightly from $E_F$, while the $d_{z^2}$ peak rapidly shifts towards $E_F$, ultimately merging with the kinks. This synchronous *d*-orbital band evolution here links $d_{z^2}$ orbital electrons to the *M*-point pairing-relevant $d_{xz/yz}$ and $d_{xy}$ orbital electrons, indicating large correlation enhancement in these *d*-orbital bands (upper inset of Fig. 2*C*).

In the corresponding zoomed-in second-derivative intensity maps (Fig. 2*D*), the red-colored regions further highlight the evolution trajectories of *d*-orbital bands (arrows) and superconducting gaps (near $E_F$). Notably, below 1.2 nA, the enhanced electronic correlation mainly contributes to the first-stage superconducting gap amplification. Within 1.2-2.2 nA, the $d_{z^2}$ peak merges with the $d_{xz/yz}$ band kink at -90 meV and subsequently merges with $d_{xy}$ band kink at -80 meV (see *SI Appendix*, Fig. S5), coinciding with the transition of stronger secondary superconducting gap amplification ($\Delta_1$ ~ 20 meV, denoted by a double-headed purple arrow). These mergers of the *d*-orbital band signatures indicate strong inter-band hybridization. Schematic reconstructed band structures at the $\Gamma$ point under two setpoint currents (Fig. 2 *E* and *F*) display the pronounced continuous energy shift of the $d_{z^2}$ band, though the hybridization and band reconstruction details (unresolved in d*I*/d*V* spectra) are omitted.

The findings here are consistently reproduced (see *SI Appendix*, Fig. S3 and S6). The absence of such peak energy shifts on the Ag surface (see *SI Appendix*, Fig. S7) further confirms the unique tip-induced behavior on FeSe/STO. Although the critical tunneling current or the relative tip-sample distance required to observe the secondary gap amplification may vary across measurements with different tips and samples (Fig. 2 and see *SI Appendix*, Fig. S6), the causal link between *d*-orbital band hybridization and the emergence of secondary gap amplification remains universally consistent in all experimental configurations. Therefore, the combined electronic correlation and band engineering (via hybridization) is demonstrated to be a viable strategy for further enhancement of superconductivity, serving as a complementary approach to carrier doping and electron-phonon coupling mechanisms.

**Repulsive tip-sample interaction**

To provide insights into the microscopic mechanism behind the tip-induced superconductivity enhancement, we show the tip-induced non-monotonic $d_{z^2}$ band shift at larger distances which mirrors the tip-sample force-distance relation (Fig. 3).

Figure 3*A* shows d*I*/d*V* spectra from 400 pA to 80 nA at 500 mV, acquired at larger tip-sample distances than those in Fig. 2 (see *SI Appendix*, Fig. S3), which are not reachable with the 30 mV data. The $d_{z^2}$ band energy exhibits a non-monotonic evolution: it initially shifts away from $E_F$, then reverses and moves sharply toward $E_F$ with increasing current. This trend is highlighted in Fig. 3*B* with the reversal point marked (also reproduced from a different sample in *SI Appendix*, Fig. S3 and S8), and shows good qualitative agreement with the schematic tip-sample force-distance profile adapted from characteristic atomic force microscopy (AFM) measurements (39) in Fig. 3*C*, where the comparison focuses on the reversal point of the curve. This correlation suggests a tip-induced in-plane lattice constant variation under tip-sample interaction. The magnitude of the energy shift further underscores this transition: under attraction, the shift is about 10-15 meV over 250 pm (Fig. 3), whereas under repulsion, it exceeds ~200 meV over a similar range (Fig. 2 and *SI Appendix*, Fig. S8), an order of magnitude larger, highlighting the dominance of repulsive interactions at closer distances (Fig. 2). This strong repulsion is further corroborated by tip manipulation of subsurface Fe-vacancy (*SI Appendix*, Fig. S9-S10).

The non-monotonic energy shift of the $d_{z^2}$ band, which mirrors the tip-sample force-distance profile, together with direct manipulation of subsurface Fe-vacancy, identifies repulsive tip-sample interaction as the mechanical origin of the perturbation. This repulsion induces in-plane lattice expansion, and given the established sensitivity of the $d_{z^2}$ band to in-plane strain, selectively drives its pronounced upward shift (16,30-34).

In intact FeSe lattice, the approaching tip exerts a progressively stronger repulsive interaction on Se/Fe atoms (particularly those directly beneath the tip apex), resulting in continuous in-plane enlargement and out-of-plane reduction of the interatomic distances (40) that effectively imposes a local in-plane tensile strain on FeSe monolayer. The inherent tensile strain imposed by the STO substrate already induces a foundational lattice expansion on the FeSe monolayer, and the pronounced tip-sample interaction synergistically amplifies the lattice expansion (Fig. 4*A*). While the absolute value of lattice expansion is challenging to measure directly under our experimental geometry (see *SI Appendix*, Note S2 and Fig. S11-S15 for details), the observed shift of the $d_{z^2}$ band provides a consistent indicator. Its movement aligns with previous ARPES studies (16), which report a ~50 meV shift of the $d_{z^2}$ band toward $E_F$ as the in-plane lattice constant expands from 3.90 Å to 3.99 Å (see *SI Appendix*, Fig. S11). In the two-dimensional FeSe layer, the tip induces such lattice expansion over a region sufficiently large and uniform to preserve a well-defined local band structure (*SI Appendix*, Note S2).

**Dual superconductivity enhancement via correlation and orbital synergy**

Lattice expansion enhances electronic correlations in FeSe, stabilizing frustrated spin fluctuations (41,42) and generating striped charge order phases (25-27,43,44). Without sufficient carrier doping, such correlation enhancement suppresses superconductivity while reinforcing charge/spin order dominance (26,27). In monolayer FeSe/$SrTiO_3$, however, the STO substrate naturally provides sufficient doping, allowing tip-induced correlation enhancement to cooperatively boost superconductivity, which constitutes the origin of first-stage gap amplification, with $d_{z^2}$ band shift reflecting electronic correlation enhancement.

Remarkably, the secondary gap amplification emerges as the $d_{z^2}$ band hybridizes with the flat $d_{xy}$ bands at -80 meV, highlighting the critical role of orbital-selective band reconstruction. This phenomenon aligns with two key electronic attributes of FeSe: (i) the orbital composition of electron-like bands crossing $E_F$ at the *M* point [including $d_{xy}$/$d_{xz}$ components (11,37), Fig. 2*B*], and (ii) the $d_{xy}$ band's larger effective mass (34,37) and dominant role in local spin susceptibility (45). These factors collectively drive the observed superconductivity enhancement through *d*-orbital band reconstruction and emergent local spin fluctuations, suggesting a potential role of both incipient bands and spin fluctuations in promoting superconducting electron pairing (46). We clarify that while the observed spectral evolution corresponds to $d_{z^2}$-$d_{xy}$ band merging at the *Γ* point, the essential band reconstruction of $d_{xy}$ orbital band is also expected to occur near the *M* point, where these pairing-relevant *d*-bands cross $E_F$. The precise origins, however, remain elusive due to limited *k*-space resolution in band structure characterization. It is unclear whether the $d_{z^2}$ orbital becomes directly involved in pairing at the *M* point during this process, despite its absence in pristine FeSe/STO (*SI Appendix*, Fig. S4). Notably, the involvement of the $d_{z^2}$ bands in Cooper pairing has been largely unexplored in iron-based superconductors, as they typically reside hundreds of meV below $E_F$. Our work achieves tunability of the $d_{z^2}$ band position (from -270 meV to -80 meV), enabling its hybridization with the $d_{xy}$/$d_{xz}$ bands near $E_F$ and leading to reconstruction of band structure, which results in enhanced superconductivity. This highlights the need for further development of pairing theories incorporating the $d_{z^2}$ orbital.

The significant roles of electronic correlation and band reconstruction in enhancing superconductivity are further corroborated by comparing the Fermi wave vector ($k_F$) values under different setpoint currents. We extract the apparent $k_{F,app}$ from spatial oscillation of Yu-Shiba-Rusinov (YSR) bound state induced by a Fe-vacancy (47) (Fig. 4 *B*-*E* and *SI Appendix*, Fig. S16). Crucially, the unchanged $k_{F,app}$ before and after the secondary gap amplification (Fig. 4 *D* and *F*) corresponds to an identical carrier density per unit cell based on the carrier density relation $n_s a^2 = (k_F a)^2/2\pi$ (*SI Appendix*, Note S3 and Fig. S17-S18). This confirms that no additional charge transfer occurs during tip approach, which excludes doping variations as the cause and strengthens band reconstruction and enhanced electronic correlation as the dominant mechanism for gap enhancement.

In addition, the pronounced tip-sample interaction can reduce the FeSe/STO interfacial distance and strengthen the interfacial interaction. This enhanced electron-phonon coupling is realized without any substrate modifications that synergizes with the enhanced electronic correlations, collectively amplifying the superconductivity.

**Discussion**

Finally, we address why tip-induced superconductivity enhancement is pronounced in FeSe/STO, yet rarely reported in other single-crystal systems. Although STM-tip modifications of YSR states are previously reported (48-50), they rarely perturb superconducting gaps. The two-dimensional nature of monolayer FeSe enables strain-sensitive structural adaptability under tip pressure, directly triggering the pronounced $d_{z^2}$ band shift and its hybridization with the bands near $E_F$. On the other hand, as mentioned before, electron doping from the STO substrate counterbalances enhanced electronic correlations, without triggering other competing correlated electronic orders (25-27,43,44,51), and enabling significant superconductivity amplification. Furthermore, our result relies on synchronous tracking of the gap and high-energy *d*-orbital band signatures over a wide range of setpoint currents, thus providing a perspective for understanding the superconductivity.

By harnessing strain-mediated energy band reconstruction to amplify superconductivity and suppressing competing orders through substrate-enabled doping, we not only reveal the $d_{z^2}$ band's influence to Cooper pairing (previously unexplored), but also demonstrate a transformative framework for engineering quantum phases. This approach could extend to diverse two-dimensional multiband superconductors, where atomic-scale lattice control may stabilize hidden correlated states. Future studies could explore multi-orbital correlation models combined with strain engineering, particularly in systems where orbital hybridization governs emergent phenomena (e.g., electronic topology). The interplay of strain, doping, and orbital selectivity opens new pathways to tailor superconducting phases in atomically designed heterostructures. Such advances offer routes to optimize quantum devices while deepening our understanding of how dimensionality, correlations, and topology intertwine in quantum materials.

## Materials and Methods

Our experiments were conducted on two independent STM-MBE systems: (i) home-built ultra-high vacuum (UHV) *in-situ* transport-STM-MBE system, (ii) commercial UHV *in-situ* STM-MBE system (Unisoku-1300).

## Monolayer FeSe/STO samples preparation

High-quality monolayer FeSe films were prepared by MBE. The Nb-doped (0.05 wt%) $SrTiO_3$(001) substrates (Shinkosha) were degassed in UHV chamber (base pressure is better than $3 \times 10^{-10}$ Torr) at 600 °C for several hours and subsequently annealed at 1200 °C for 20 min to obtain the $TiO_2$-terminated surface. Then, high-purity Fe (99.995%) and Se (99.9999%) sources were coevaporated by two Knudsen cells to deposit monolayer FeSe films on the $SrTiO_3$ substrates. During deposition, the substrates were maintained at either 450 °C (i) or 430 °C (ii) via direct current heating. The as-grown films were subjected to stepwise annealing from 480 to 520 °C for hours to optimize sample quality. These samples underwent additional postannealing at 560 °C (i) or 550 °C (ii) for several tens of minutes to fully optimize the superconducting gap. Finally, the samples were transferred to the *in-situ* low-temperature STM for studies.

## STM measurements

*In-situ* STM measurements were performed at 4.2 K. To ensure the reproducibility, polycrystalline Pt-Ir STM tips were subjected to an *in-situ* regeneration procedure before each independent measurement. First, the tips were heated by electron-beam irradiation to temperatures above 600 °C for several minutes to completely remove residual apex structures of tips. Then, the tips were recalibrated on the freshly prepared Ag(111) surface.

The tip-sample distance was precisely regulated by tuning the tunneling current at a fixed setpoint bias voltage under feedback control. The Proportional-Integral parameters, which were optimized for the specific STM, ensured stable tip-sample distance at the designated setpoint current. Subsequent STS measurements were performed at these stabilized tip-sample distances (determined by setpoint bias and current values) by sweeping bias with the feedback loop disabled. STS data were acquired by standard lock-in technique with a reference oscillation frequency of 973.0 Hz.

## Manipulation of Fe-vacancy

The manipulation of Fe-vacancy was performed by approaching the STM tip to the specific lattice position (*SI Appendix*, Fig. S9*A*) while maintaining a fixed setpoint bias of 15 mV (in-gap) with repeatedly varying the tunneling current from 4.5 nA to 9.5 nA.

## Data availability

All study data are included in the article and/or *SI Appendix*.

**Acknowledgments**

We thank P. J. Hirschfeld, J. P. Hu, Y. Zhang, and H. Yao for helpful discussions. The experimental work was supported by the National Natural Science Foundation (Grants No. 92365201, No. 52388201, No. 11427903), the National Key R&D Program of China (Grants No. 2022YFA1403100), and the Innovation Program for Quantum Science and Technology (2021ZD0302402).

## Figures and Tables

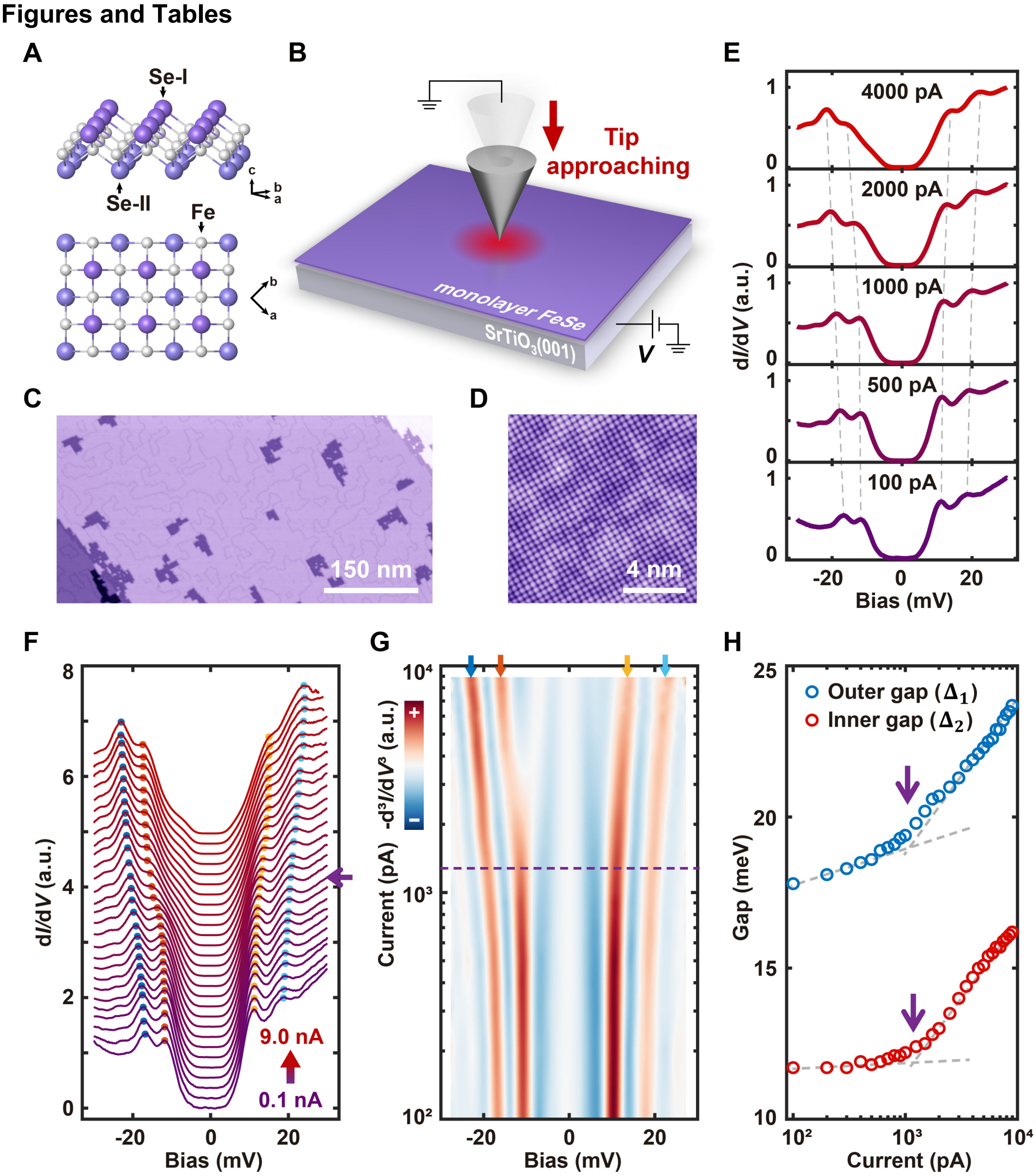

**Figure 1.** Tip-driven dual enhancement of superconducting gap in monolayer FeSe/STO. (*A*) Crystal structure of FeSe. (*B*) Schematic of the experiment. (*C*) Large-scale STM topography of monolayer FeSe/STO (bias voltage $V$ = 2.0 V, tunneling current $I_t$ = 20 pA). (*D*) Atomically resolved STM topography of topmost Se-layer (Se-I) ($V$ = 30 mV, $I_t$ = 8 nA). (*E*) Tunneling current-dependent d$I$/d$V$ spectra, exhibiting enlarged superconducting gaps under decreasing tip-sample distance. Dashed lines trace the energy shifts of coherence peaks. (*F*) Normalized d$I$/d$V$ spectra of superconducting gap (100 pA to 9 nA; vertical offset for clarity). Colored dots highlight the coherence peaks. (*G*) Second-derivative [$-d^2(dI/dV)/dV^2$, i.e., $-d^3I/dV^3$] intensity maps of (*F*) (logarithmic current scale), highlighting the evolution of the coherence peaks marked by arrows. Purple dashed line indicates a transition of gap evolution [defined in (*H*)]. (*H*) Current-modulated superconducting gap evolution (logarithmic scale). Dashed lines track the gap evolutions; purple arrows mark the transition. Gap energies are extracted as the mean of the asymmetric positive and negative coherence peaks. Setpoint in (*E-G*), $V$ = 30 mV.

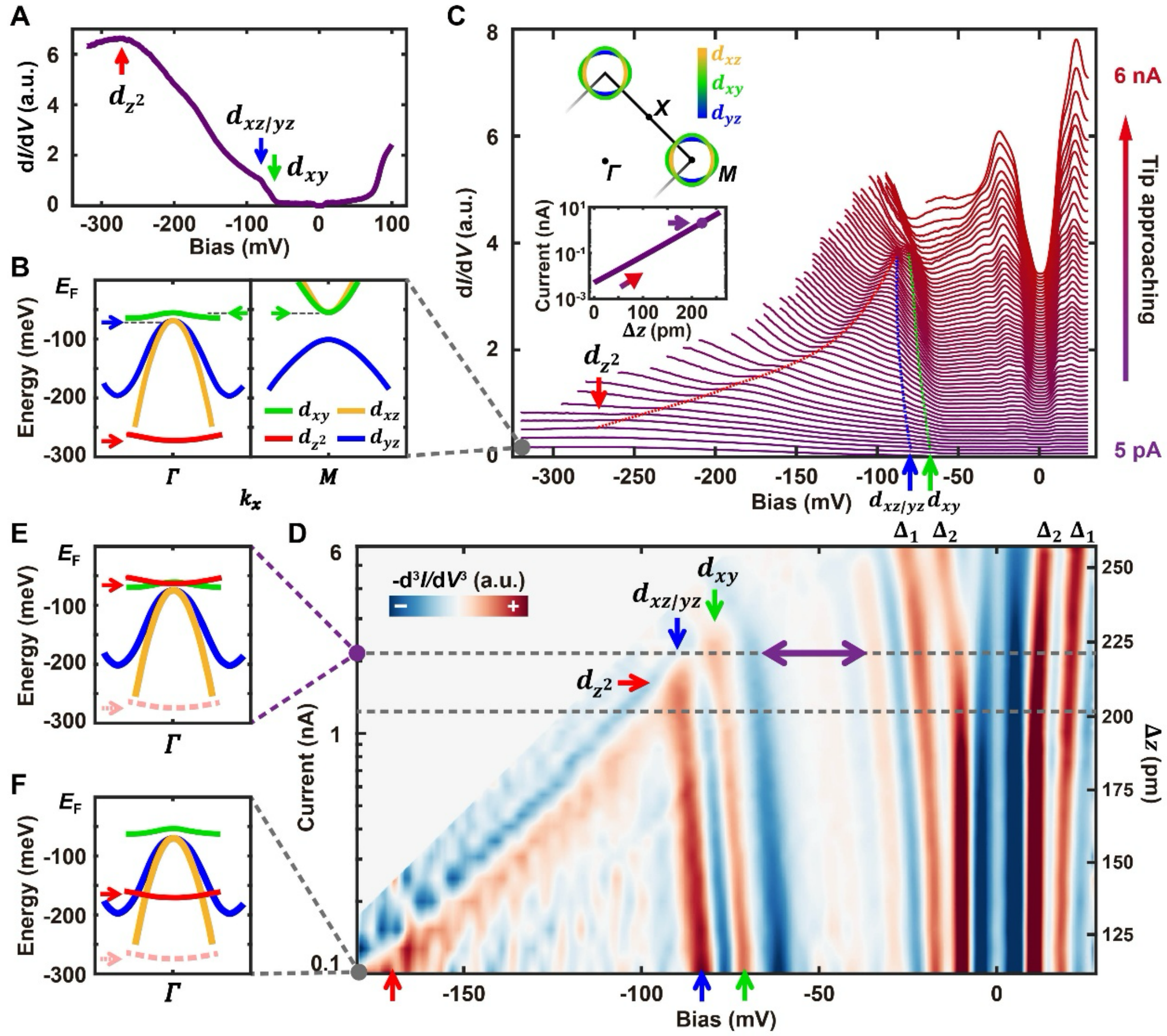

**Figure 2.** Tip-driven *d*-orbital band reconstruction and superconducting gap dual amplification in monolayer FeSe/STO. (*A*) Wide-bias-range d*I*/d*V* spectrum ($V$ = 200 mV, $I_t$ = 2 nA). Red arrow marks the $d_{z^2}$ band peak. Blue and green arrows indicate two kinks attributed to the $d_{xz/yz}$ and $d_{xy}$ bands. Color representations remain consistent in subsequent figures. (*B*) Schematic band structure and orbital characters of monolayer FeSe/STO [adapted from ref. (11,37)]. (*C*) Current-dependent d*I*/d*V* spectra ($V$ = 30 mV, $I_t$ from 5 pA to 6 nA; vertical offsets for clarity), revealing the evolutions of the *d*-orbital bands and superconducting gaps, simultaneously. Dashed lines track the band energy shifts. Upper inset: Fermi surface schematic [adapted from ref. (38)]. Lower inset: Linear response of logarithmic tunneling current with reducing tip-sample distance ($\Delta z$ = 0 is defined at $V$ = 30 mV, $I_t$ = 5 pA). The purple arrow marks the transition point of gap evolution. (*D*) Second-derivative intensity maps of data in (*C*) (100 pA to 6 nA, logarithmic current scale), the corresponding $\Delta z$ values are displayed on the right axis. (*E-F*) Schematic band structures at $\Gamma$ point under two setpoint currents: 2.2 nA for (*E*), and 100 pA for (*F*). The $d_{z^2}$ band shifts upwards to -80 meV and -180 meV, respectively.

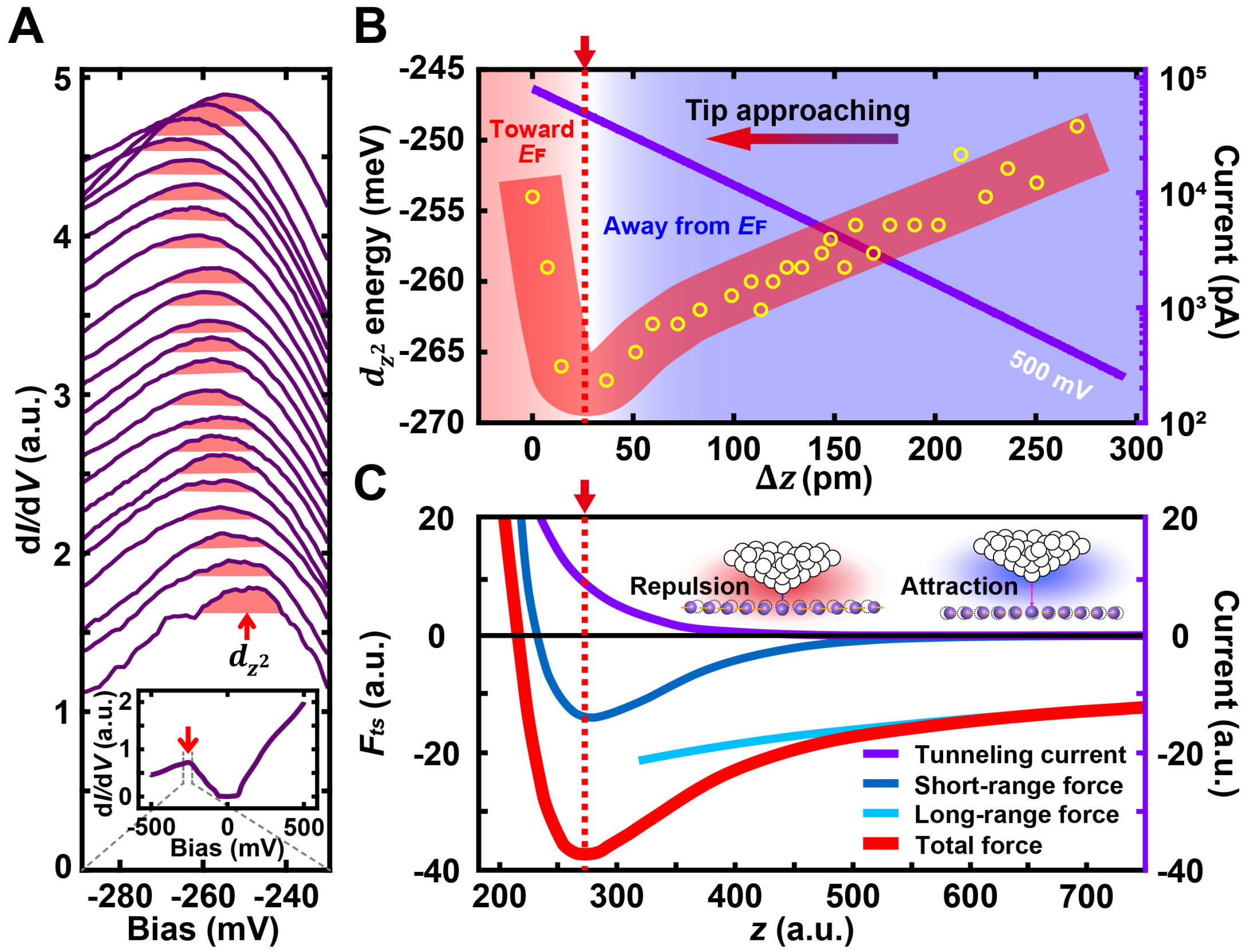

**Figure 3.** Tip-driven non-monotonic $d_{z}{}^{2}$ band shift mirrors tip-sample force-distance relation. (*A*) Normalized d*I*/d*V* spectra ($V$ = 500 mV, $I_t$ from 400 pA to 80 nA; vertical offsets). Red shades highlight the $d_{z}{}^{2}$ band peaks on each spectrum, outlining the energy shift. Inset: Wide-bias-range d*I*/d*V* spectrum ($V$ = 500 mV, $I_t$ = 1.25 nA) with the analyzed energy range marked. (*B*) $d_{z}{}^{2}$ band energy (left *y*-axis, yellow circles) and setpoint current (right *y*-axis, purple line) versus tip-sample distance variation. Red dashed line marks the reversal point in the $d_{z}{}^{2}$ band evolution. (*C*) Schematic tip-sample force-distance curve [adapted from ref. (39)]. Red dashed line marks the position of maximum attraction defined as the reversal point of the total force curve. Insets illustrate the schematics of the attractive and repulsive interaction regimes.

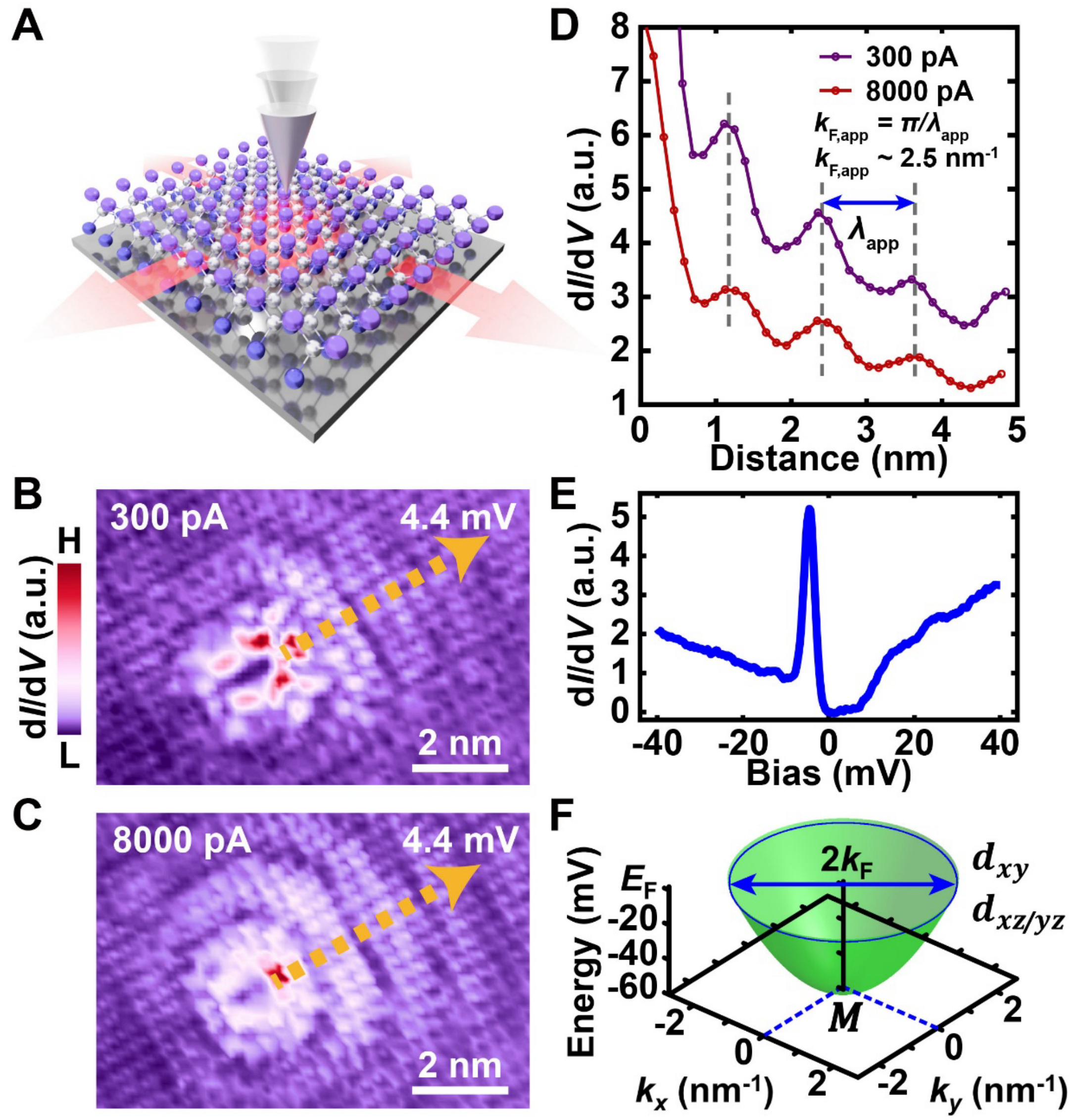

**Figure 4.** Invariant Fermi wave vectors during superconducting gap enhancement. (*A*) Tip approach generates local in-plane lattice expansion in FeSe. (*B-C*) Spatially resolved YSR bound states around an Fe-vacancy under setpoint currents of 300 pA (*B*) and 8000 pA (*C*) ($V$ = 30 mV). Dashed arrows indicate the direction of the DOS line cuts. (*D*) Spatial distribution of normalized DOS under the two setpoint currents along the dashed arrows (vertical offset for clarity). The YSR bound states exhibit identical apparent wavelengths under both setpoint currents. (*E*) d$I$/d$V$ spectrum measured at the Fe-vacancy ($V$ = 200 mV, $I_t$ = 2 nA). (*F*) Schematic of the electron-like pocket and $k_F$ around the $M$ point in FeSe/STO.